\documentclass[10pt,twocolumn,letterpaper]{article}
\usepackage{dcolumn}
\usepackage{algpseudocode}
\usepackage{algorithmicx}
\usepackage{algorithm}
\usepackage{cvpr}
\usepackage{times}
\usepackage{epsfig}
\usepackage{graphicx}
\usepackage{amsmath}
\usepackage{amssymb}
\usepackage{amsthm}

\usepackage{caption}

\usepackage{todonotes}

\def\ct{\hbox{$T_{C}$}}
\def\st{\hbox{$T_{S}$}}
\def\mid{\textit{MID}}
\def\uid{\textit{UID}}
\def\mt{\hbox{$T_{M}$}}

\long\def\comment#1{}

\newcommand*\xor{\mathbin{\oplus}}
\makeatletter
\def\tightmath{
\abovedisplayskip=4pt plus 2pt minus 1pt 
\abovedisplayshortskip=2pt plus 1pt minus 1pt 
\belowdisplayskip=4pt plus 2pt minus 1pt 
\belowdisplayshortskip=2pt plus 1pt minus 1pt }
\def\crushmath{
\abovedisplayskip=1pt plus 1pt minus 2pt 
\abovedisplayshortskip=1pt plus 1pt minus 2pt 
\belowdisplayskip=1pt plus 1pt minus 2pt 
\belowdisplayshortskip=1pt plus 1pt minus 2pt }
\tightmath
\crushmath

\makeatother

\usepackage[pagebackref=true,breaklinks=true,letterpaper=true,colorlinks,bookmarks=false]{hyperref}

\cvprfinalcopy 

\def\cvprPaperID{1272} 

\ifcvprfinal\fi
\begin{document}

\title{CALIPER: Continuous Authentication Layered with Integrated PKI Encoding Recognition}

\author{
Ethan M. Rudd and Terrance E. Boult\\
University of Colorado at Colorado Springs\\
Vision and Security Technology (VAST) Lab\\
\{erudd,tboult\}@\url{vast.uccs.edu}
}

\maketitle
\thispagestyle{empty}


\begin{abstract}
Architectures relying on continuous authentication require a secure way to challenge the user's identity without trusting that the Continuous Authentication Subsystem (CAS) has not been compromised, i.e., that the response to the layer which manages service/application access is not fake. In this paper, we introduce the CALIPER protocol, in which a separate Continuous Access Verification Entity (CAVE) directly challenges the user's identity in a continuous authentication regime. Instead of simply returning authentication probabilities or confidence scores, CALIPER's CAS uses live hard and soft biometric samples from the user to extract a cryptographic private key embedded in a challenge posed by the CAVE. The CAS then uses this key to sign a response to the CAVE. 
CALIPER supports multiple modalities, key lengths, and security levels and can be applied in two scenarios: One where the CAS must authenticate its user to a CAVE running on a remote server (device-server) for access to remote application data, and another where the CAS must authenticate its user to a locally running trusted computing module (TCM) for access to local application data (device-TCM). We further demonstrate that CALIPER can leverage device hardware resources to enable privacy and security even when the device's kernel is compromised, and we show how this \textit{authentication protocol} can even be expanded to obfuscate direct kernel object manipulation (DKOM) malwares.
\end{abstract}

\section{Introduction}
\label{sec:intro}

Two fundamental problems exist with conventional password, token, and biometric authentication schemes. 
The first problem is one of \textit{infrequent authentication}. 
Assuming that a legitimate user performed the initial authentication, there is no validation mechanism to ensure that the same user that logged in is the same user using the device minutes or hours later. 
Such authentication schemes guarantee only that the current user of the device possessed a password, biometric, or security token at the time of login. 
This guarantee does nothing to address post-login risk: If Bob steals Alice's phone while Alice is still logged on to her bank account, Bob can do catastrophic harm to Alice, regardless of how many questions the bank asked or the number of factors of authentication used during the initial login.

The second problem is one of \textit{remote trust}.
 Remote services must trust the authentication subsystem: They have no independent way to challenge a user's identity. 
If Bob steals Alice's authentication credentials, Bob will still be authenticated by Alice's bank, since the bank can only establish the veracity or lack thereof of the login credentials. 
Bob can continue to login until Alice or the bank notice and/or change Alice's password.
Note that device authentication methods by themselves, in which the user authenticates to the device and the device authenticates to a remote entity still do not solve the remote trust problem. 
Even under the idealized fictional assumption of perfect device authentication, i.e., the remote entity only knows that \textit{the actual device's} user authentication module ``said'' that the user was valid. It has no independent way to challenge the user. If Bob has stolen Alice's device and password, he poses a security risk, even under this naive fictional assumption of perfect device authentication.

These two shortcomings in conventional authentication schemes can allow an adversary to gain access to the device post-login, which at the very least compromises the session. Further, they may allow an adversary to steal or spoof login credentials for later use at the adversary's discretion, potentially compromising confidentiality, integrity, and availability of all data to which the legitimate user of the device has access. 

Recent advances in continuous authentication offer promising solutions to the infrequent authentication problem. In continuous authentication, a device continuously obtains hard and soft biometrics in a manner transparent to the user so that local or remote services may continuously determine whether to maintain the user's authentication credentials as legitimate or to de-authenticate the user. Moreover, continuous authentication offers the appeal of enhancing user experience by reducing explicit password prompts.
To our knowledge, however, continuous authentication architectures to date \textit{focus solely on authentication confidence}. This does not address the remote trust problem: Even if authentication subsystems use biometrics that provide strong authentication confidence, it is insufficient for the continuous authentication layers to return the probabilities of valid users, because other local layers or remote services then have to trust that the response was not fake. Whether the client device is physically compromised via theft or remotely compromised via malware, an adversary need only change one bit to defeat such weak security.

To overcome the limitations of conventional authentication systems, we present a protocol to harden continuous authentication implementations against scenarios in which the local or remote Continuous Access Verification Entity (CAVE) cannot trust that the Continuous Authentication Subsystem (CAS) client has not been compromised by theft or malware on the device.
Henceforth, we refer to this authentication protocol CALIPER -- \textit{Continuous Authentication Layered with Integrated PKI Encoding Recognition}. CALIPER is a continuous authentication protocol, where continuous hard (e.g., face, speech) and soft (e.g., keystroke data, application data, resource usage data) biometric data are collected using existing device I/O API infrastructure and are then transformed into an intermediate classifier / feature space representation for use as basic blocks in a biocryptographic ensemble. Unlike traditional biometric verification schemes or other continuous authentication approaches, the result is a system that leverages a PKI encoding to support challenge-response key exchange so that other local or remote services do not have to trust only the output of the continuous authentication subsystem (CAS); Instead they can challenge the CAS directly about a given user and even use biocryotpographics~\cite{scheirer2013beyond} to exchange/manage public keys. The contributions of this paper are as follows:
\begin{enumerate}
\item We introduce CALIPER, the first authentication protocol that addresses both \textit{infrequent authentication} and \textit{remote trust} problems with conventional authentication schemes. CALIPER generalizes password-combined, one-time, single-modality remote \textit{vaulted verification} (VV) protocols to a \textit{continuous, multiple-modality, biocryptographic key exchange/renewal protocol}. CALIPER's basic blocks are not dependent on any one classifier type or even a machine learning classifier in the traditional sense, which allows CALIPER to form challenges based on CAS device specifications too, becoming a \textit{unified 3-factor authentication protocol}.

\item We present different scenarios for applying CALIPER as an \textit{access control protocol}: a \textit{device-server} scenario, where a CAS residing on a client device must authenticate its user to a CAVE running on a remote server, and a \textit{device-TCM} scenario, where a CAS must authenticate a user to a CAVE resident on a trusted computing module (TCM), e.g., a TPM, SIM card, or GPU running on the same local device, in order to run an application.

\item Although CALIPER is an authentication protocol, its flexibility allows us to apply it to other problems in computer security. We demonstrate how the protocol can be used to maintain security even in the event of OS kernel compromise, and even obfuscate attacks through \textit{Address Space Layout Personalization} (ASLP), a novel approach to address space layout randomization (ASLR), whose security does not depend on an intact kernel. 
\end{enumerate}
\section{Background}
\label{sec:background}
The CALIPER protocol combines the concepts of continuous authentication and biocryptographic challenge-response.
The earliest work on continuous authentication systems dates back to the early 1990's, in which Leggett et al. proposed using keystroke characteristics as a ``dynamic identity verifier'' to overcome the static nature of session passwords: ``Normally, the user is asked for the password at log-in time and the system assumes that the user is the same person until log-off time''~\cite{leggett1991dynamic}. 
The term “continuous authentication” in the biometric sense dates back to 1995, again applied to keyboard dynamics~\cite{shepherd1995continuous}. 
Since then, various research has been published on applying continuous authentication across many modalities, e.g.,~\cite{azzini2008fuzzy,feng2012continuous,frank2013touchalytics,guennoun2009continuous,klosterman2000secure,liu2009optimal,jorgensen2011mouse}. 

Recently, continuous authentication has gained increased interest in the mobile domain, largely due to the explosion of the mobile device market, the increasing number of sensors available on mobile devices, the increasing need for mobile security, and the increased accuracy and decreased processing costs of applying machine learning algorithms. 
Large companies are now actively pursuing continuous authentication frameworks for mobile devices, for example, Google Inc. recently announced Project Abacus~\cite{project_abacus}, a fused multi-modal continuous authentication framework for Android devices. 
However, no continuous authentication framework to date to our knowledge addresses the problem of CAS compromise on the local device, but instead, these frameworks implicitly trust that the CAS has not been compromised and that there is not, e.g., a rootkit hooking CAS binaries. We contend that this implicit trust is the weak link in the security of most proposed, prototyped, and fielded continuous authentication systems.

\comment{ CALIPER guarantees security and privacy via an extension of the Index-Table Vaulted Verification (IVV) protocol~\cite{johnson2013vaulted}, which is in turn an enhancement of the Vaulted Verification (VV) protocol.  VV was originally formulated in~\cite{wilber2012secure} for face biometrics to address problems in popular biometric authentication and template protection schemes, including template privacy/security vulnerabilities in the event of server compromise, man-in-the-middle attacks, replay attacks, blended substitution attacks, and non-revocability of biometric models/tokens.}

CALIPER guarantees security and privacy via an extension of the Vaulted Verification protocol proposed by Wilber et al. in~\cite{wilber2012secure}.  
VV was originally proposed as an extension of fuzzy vaults ~\cite{juels2006fuzzy} to address the numerous security problems with fuzzy vaults and fuzzy extractors~\cite{dodis2004fuzzy} that were reported by Scheirer et al. in~\cite{scheirer2007cracking}. These problems include template privacy/security vulnerabilities in the event of server compromise, man-in-the-middle attacks, replay attacks, blended substitution attacks, and non-revocability of biometric models/tokens~\cite{scheirer2007cracking}. Johnson presents over 60 pages of analysis on the security of the VV protocol in~\cite{johnson2014privacy}. This analysis indicates that the protocol, under proper implementation, is secure against such attacks.

The VV protocol is a client-server protocol that can be summarized as follows: During enrollment, the client takes real biometric samples and generates chaff samples. 
Models for both real and chaff are created and (real, chaff) model pairs, each pair encrypted with the client key, all pairs encrypted with the server key, along with corresponding ground truth, are sent to the server. 
The client then securely wipes all data from memory. 
During authentication, the server generates a random key with which it changes the order of each (real, chaff) pair depending on whether the corresponding key element is 1 or 0 respectively. 
The server then sends a challenge to the client, consisting of pairs swapped according to the key. 
Upon receiving the pairs, the client, using live biometric samples in conjunction with the challenge models, responds with a guess of the key. 
Upon receiving the client's guess, the server checks against the actual key. 
If enough bits are correct for the security policy in question, the server authenticates the client. 
VV also uses nonces to protect against replay attacks.

There have been several variations of the VV protocol over the years, including extensions to iris~\cite{wilber2012privv}, fingerprint~\cite{alzahrani2014remote}, and voice~\cite{johnson2013secure,johnson2013voice,johnson2013vaulted} biometrics. The latter extensions incorporated public key cryptography, and reduced communication overhead by using \textit{index tables}. 
All of these protocols, however, use VV strictly as a one-time, single modality remote authentication mechanism to be combined with password authentication. Therefore, \textit{none of them} can be incorporated into continuous authentication solutions, which necessarily must support multiple modalities~\cite{sim2007continuous}, for sufficient authentication capability and a usable experience in the event of failover of one or more modality types. 
This claim is not merely theoretical: commercial development efforts~\cite{project_abacus} have leveraged several modalities in their continuous authentication algorithms.
CALIPER extends previous VV efforts to a \textit{continuous, multiple-modality, biocryptographic key exchange/renewal protocol}. 

Related to but not to be confused with the \textit{remote trust} authentication problem discussed in Sec.~\ref{sec:intro} is the problem of \textit{remote entrusting}~\cite{scandariato2008application,aussel2009smart,ceccato2008remote}, a more general computer security problem in which an application running on an untrusted device requests resources from a remote entity. Solving the \textit{remote trust} authentication problem does not solve the \textit{remote entrusting} problem, because establishing that a user is authentic does not ensure that the device has not been compromised by malware. If a rootkit has hooked the device while a legitimately authenticated user is viewing sensitive data, that sensitive data can still be compromised at no fault of the user authentication subsystem.
Remote entrusting is an extremely difficult problem to solve because of the problems presented by modern malwares, especially those which hijack control flow of either legitimate applications or the kernel itself~\cite{rudd2016survey}. \textit{Solving} the remote entrusting problem is well beyond the scope of this paper, if it is even possible to begin with. We can, however, \textit{mitigate} the problem by extending CALIPER \textit{beyond} an access control protocol. We present this extension in Sec.~\ref{sec:aslp}. 

\section{Protocol}
\label{sec:caliper_protocol}
In this section we discuss the mechanics of the CALIPER protocol. The notation used in this protocol is summarized in Table~\ref{tab:notation}. The protocol has two stages: enrollment and verification. We present the enrollment stage in Sec.~\ref{sec:enrollment} and the verification stage in Sec.\ref{sec:verification}. 

\begin{table}[H]
\begin{center}
\begin{tabular}{ | l | p{6cm} |}
\hline
{\bf Symbol} & {\bf Definition} \\
\hline
\small $(K_{pu},K_{pr})$ & \small Public and private key pair generated by the CAS. \\
\hline
\small $C(K_{pr})$ & \small $K_{pr}$ under an ECC encoding. \\
\hline
\small $\hat{K}$ & \small $C(K_{pr}) \xor R$\\
\hline
\small $K_{perm}$ & \small Permutation key used by the server to generate challenges. \\
\hline
\small \mt & \small Model table resident on CAS.\\
\hline
\small \ct & \small Client table residing on the CAS.\\
\hline
\small \st & \small Server table residing on the CAVE.\\
\hline
\small \hbox{$h1$} & \small Hash index into \mt.\\
\hline
\small \hbox{$h2$} & \small Hash index into \ct.\\
\hline
\small \hbox{$h3$} & \small Hash of retrieved $K_{pr}$, shift amounts, and $n2$ sent as a response to the CAVE.\\
\hline
\small \hbox{$n1$} & \small Nonce 1.\\
\hline
\small \hbox{$n2$} & \small Nonce 2.\\
\hline
\small $R$ & \small A random pad.\\
\hline
\small \uid & \small User identifier.\\
\hline
\small \mid & \small Modality identifier.\\
\hline
\small $H$ & \small hash(\st,$K_{pu}$,\uid).\\
\hline
\small $N$ & \small Number of choices per row ($\geq 2$) in \st.\\
\hline
\small $M$ & \small Number of rows in \st. Same as the number of challenges available at a particular time.\\
\hline
\small $i$ & \small Column index in \st; $i \in \{1,\hdots,N\}$.\\
\hline
\small $C$ & \small Challenge.\\
\hline
\small \hbox{$RE$} & \small Response.\\
\hline
\end{tabular}
\end{center}
\caption{\label{tab:notation} \small Notation used in the CALIPER protocol.}
\end{table}

\subsection{Enrollment}
\label{sec:enrollment}

\begin{figure*}[!th]\begin{center} \includegraphics[width=\textwidth]{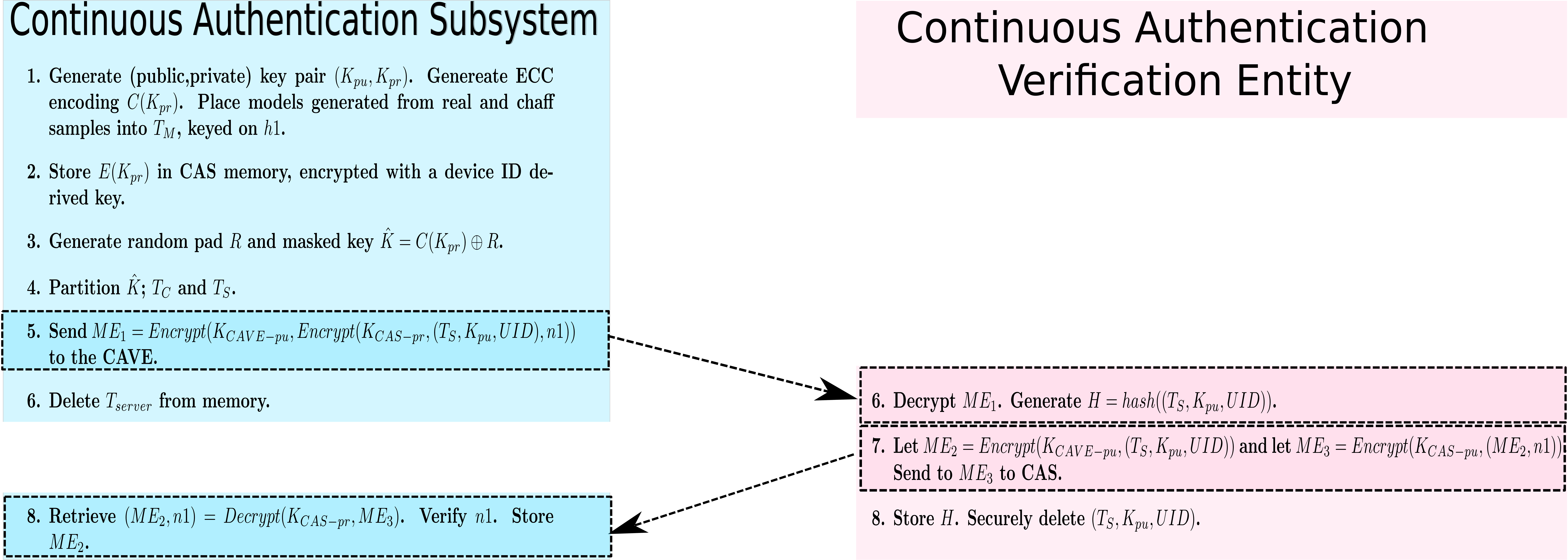}\hfil \end{center}
\caption{\small
Communication and processing tasks on the CAS and the CAVE in rough time sequence order during enrollment. The presence of multiple numbers indicates that multiple tasks occur at approximately the same time. Dashed arrows indicate communication between the CAS and the CAVE. White space designates times when a device is idle. Note that communication between CAS and CAVE is protected by an encrypted channel.}
\label{fig:enrollment_communication}
\end{figure*}

The enrollment portion of the CALIPER protocol is depicted in communication and processing steps in Fig.~\ref{fig:enrollment_communication}. Enrollment proceeds as follows: The CAS continuously polls different sensors and OS resources for data from a subset of modalities, extracting feature vectors and generating an ensemble of classifiers for both real and chaff samples. 
The classifier ensemble is then stored in a model table (\mt), in which each element is a classifier, keyed on its own hash.
The CAS then generates a (public, private) key pair $(K_{pr} , K_{pu})$, followed by an ECC representation $C(K_{pr})$.
A random pad, $R$, of the same length as $C(K_{pr})$, is then generated and a masked key, $\hat{K} = C(K_{pr})\xor R$, is created. $\hat{K}$ is then partitioned into chunks and a client table (\ct) is created. Each entry in \ct\ consists of a hash index ($h1$) into \mt, which is a direct hash of each classifier; an index $i$, which assumes values between 1 and $N$, the number of possible answers allowed to the CAS at verification time for this key fragment; $\hat{K}_i$ , which corresponds to the key if $i$ corresponds to the actual answer (real sample) and a random number otherwise (chaff sample); and finally the modality identifier (\mid). 

Each row of \ct\ is keyed on $h2$, a hash of its contents. An $M$-row server table (\st) is then created, with each row containing an entry for a real classifier of a different modality and $N - 1$ chaff entries, where $N$ is the number of choices per row, greater than or equal to 2. Each entry in \st\  contains $h2$, the hash index into \ct, the column index, $i$, and the associated segment of the random pad, $R_i$, used to retrieve the key if the column indexes a real model and another random number otherwise. Note that \st\ may have variable-length columns, but for simplicity of discussion we treat them as fixed-length. After \ct\ and \st\ are generated, $\hat{K}_i$,$C(K_{pr})$, and $K_{pr}$ are wiped from the client's memory.

The CAS then encrypts \st, along with the public key that was generated by the CAS, the \uid, and \hbox{$n1$} with the CAVE's public key and sends them to the CAVE. The CAS deletes its copy of \st. The CAVE then decrypts the incoming message, extracts \st, concatenates \st\ with \hbox{$n1$}, encrypts the pair with its public key, hashes this encrypted pair, stores the hash, sends the result back to the CAS along with \hbox{$n1$} over an encrypted channel, and securely wipes its own copy of the server table from memory. The CAS then verifies \hbox{$n1$} to ensure that the encrypted \st\ corresponds to the same communication session and stores the encrypted \st. This concludes enrollment.

\subsection{Verification}
\label{sec:verification}

\begin{figure*}[!ht]\begin{center} \includegraphics[width=\textwidth]{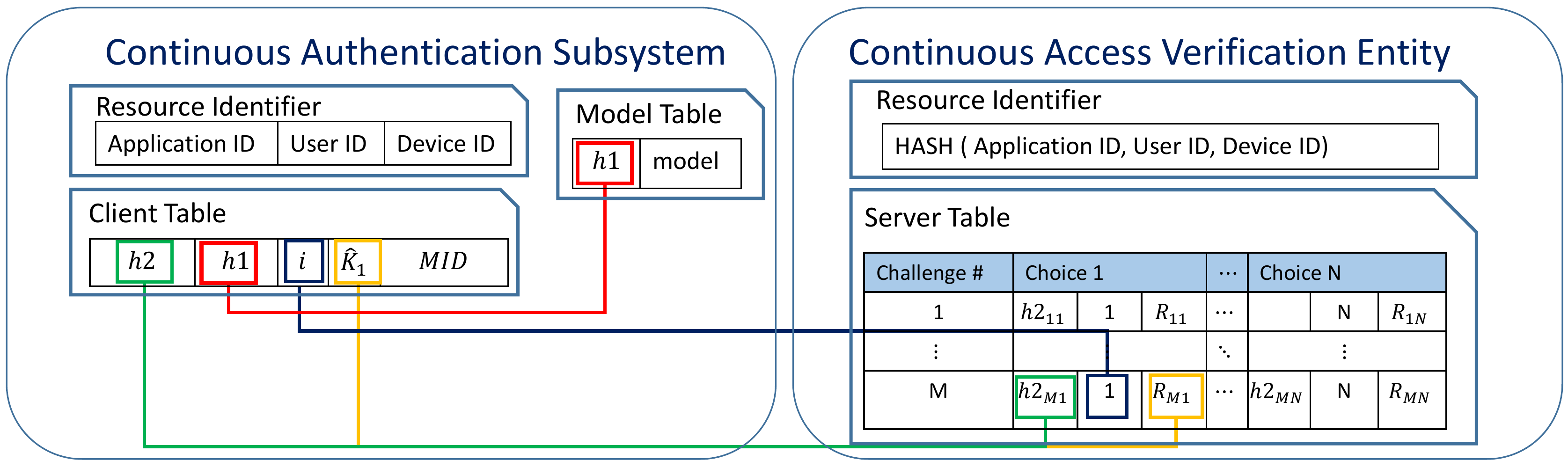}\hfil \end{center}
\caption{\small The layout of index-table resources in CAS and CAVE memory. 
Related elements across tables are connected with solid lines.
For ease of visualization, we show a rectangular server table with $M$ challenges with $N$ choices per challenge. 
In the server table, indices are made explicit on $h2$, $i$ (the choice), and $R$.
We show only one entry of the client table, omitting indices on $h1$ and $h2$. 
We index $\hat{K}$ by $i=1$ to denote the explicit correspondence with $R_{M1}$ in the server table. 
The layout depicted in this diagram is before the challenges in the server table have been circularly shifted; otherwise the correspondences would likely be broken.
Note also that the correspondences across tables are only hashes/indices. 
The CAVE sees no biometric/behavioral data; not even the type of modality.
The CAS stores only model files with no associated ground truth.
Also, the CAVE's session resource need not store user information, device information, or even names of applications.
Thus, CALIPER not only provides strong security but also significantly enhances privacy.}
\label{fig:verification_communication}
\end{figure*}

The verification portion of the CALIPER protocol is depicted in communication and processing steps in Fig.~\ref{fig:verification_communication}.
When the CAS first requests access from the CAVE or when the continuous authentication protocol requires a new session key, the CAS sends the CAVE a request for access, which includes the encrypted copy of the server table. The CAVE hashes this message and compares the generated hash against its previously stored hash to ensure that the contents of the server table have not changed. The CAVE then decrypts the message using its private key, and constructs a challenge for the CAS by first selecting a random subset of rows from the server table, then creating a vector of random integers the length of this permuted subset of rows. Each element of this vector assumes a value between 0 and $N-1$, where $N$ is the length of the corresponding row in the server table. The subset  and permutation of rows are then circularly shifted by their corresponding random element, and, with the exception of the original index, sent to the CAS. A challenge nonce \hbox{$n2$} is included in this message to protect against replay attacks.

When the CAS receives the challenge, it begins constructing the response, element by element, by first indexing into \ct\ and retrieving appropriate values by which to index into \mt.
The CAS then indexes into \mt\ to retrieve the corresponding challenge classifiers. By submitting live biometric/behavioral samples to each of the retrieved classifiers, the CAS can determine which models were generated using real data. This also allows the CAS to guess the shifts that were applied by the CAVE.
The CAS can then retrieve fragments of the encoded key by XORing the encoded key fragments from \ct\ with the associated random pad from the challenge, i.e., \hbox{$C(K_{pr})_i = \hat{K}_i \xor R_i$}. Provided that the CAS successfully retrieves enough codeword fragments, it can run error correction on $C(K_{pr})$ to recover $K_{pr}$. It can then piece together the permutation key $K_{perm}$ that was used by the server, and sign a response via the hash \hbox{$h3=hash(K_{pr},K_{perm},n2)$}.  When the CAVE decrypts the message and successfully verifies the hash it authenticates the CAS device's user. The CAS client can also include a message with a vector of its original guesses on the permutations. Depending on the number of guesses that the client got correct, the challenge unit can adjust the time intervals / key lengths required between subsequent authentications. 

\begin{figure*}[!t]\begin{center} \includegraphics[width=\textwidth]{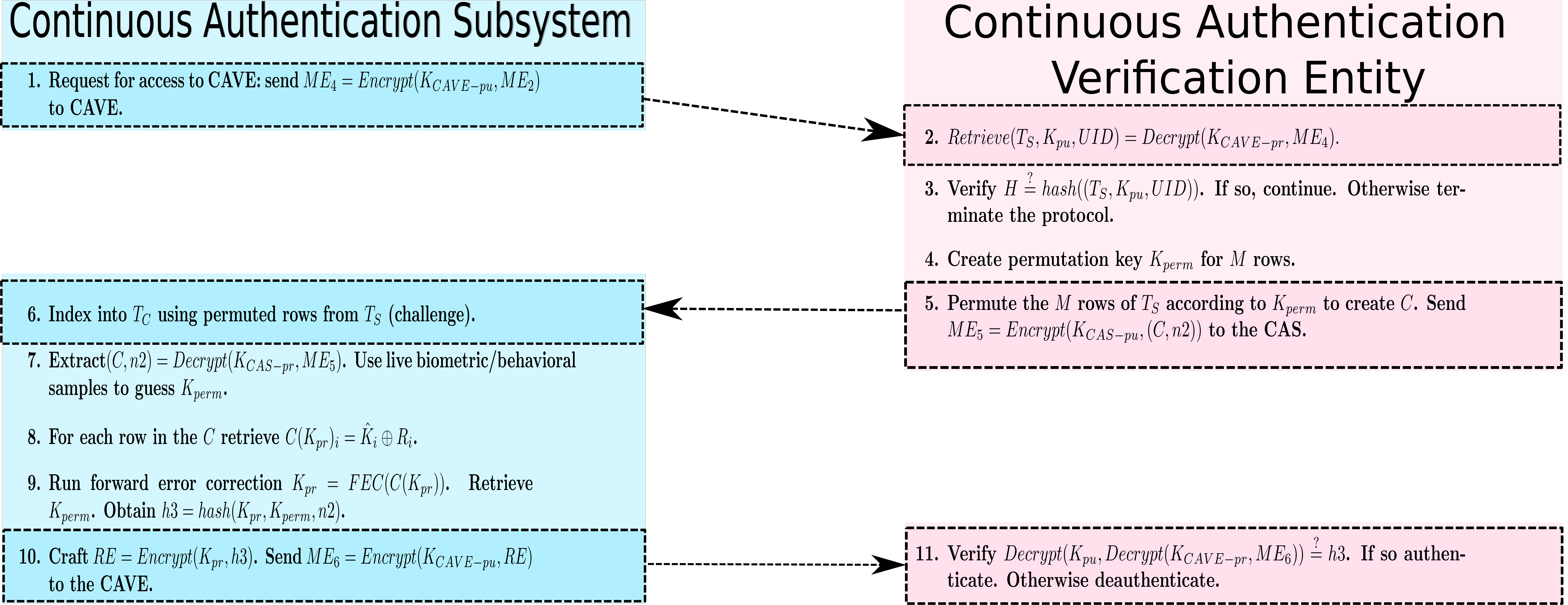}\hfil \end{center}
\caption{\small
Communication and processing tasks on the CAS and the CAVE in rough time sequence order during verification. Dashed arrows indicate communication between the CAS and the CAVE. Times during which a device is idle are indicated by white space. Note that communication between CAS and CAVE is protected by an encrypted channel.}
\label{fig:indextable}
\end{figure*}

\subsection{Implementation Considerations}

Several of the security enhancements that CALIPER provides over IVV are non-trivial to quantify: First, the extension to multiple modalities makes spoofing CALIPER more difficult than spoofing a single-modality IVV scheme. Depending on the implementation, it might be desirable for the CAVE to be able to alternate the ratios/types of modalities used, which requires placing the \mid\ in the server table. 
A second way CALIPER enhances security over conventional IVV is that it generalizes the IVV basic block to any classifier type, thus allowing more bits of security to be generated faster than under a single modality regime. Previous single modality VV schemes treat the basic block as an entity consisting of one homogeneous model type. 

CALIPER's basic blocks can consist of any type of model on which training/classification can be performed in a reasonable amount of time. For the face modality, for example, basic blocks might consist of support vector machines, while for the voice modality, basic blocks might consist of Gaussian mixture models. Interestingly, a basic block need not even correspond to a model in the conventional machine learning sense; in some cases simple comparisons of values may suffice. Let us say that we wanted another way to verify the device itself beyond conventional device authentication protocols. By using hashes of disk blocks as basic blocks in the model tables, along with the \textit{logical block addresses} (LBAs) associated with the hashes, “classification” reduces to indexing disk blocks, hashing them, and performing equality checks to determine which hashes correspond to the actual storage layout of the device. Leveraging multiple modalities and supplementing with simple comparison-style blocks allows CALIPER to have more bits of security on-demand than single modality schemes, and provides an third factor of authentication (something the user has) over the IVV protocol. Of course, some balance of modalities is necessary to provide secure authentication. 

We would also like to point out that, in implementation, the security of the protocol assumes that a separate $(K_{pu},K_{pr})$ pair is used at each verification step. Otherwise malware on the device could simply snoop $K_{pr}$, thus obviating challenge-response.

\section{Applications}

CALIPER offers a secure continuous authentication model which can be mixed across many devices of heterogeneous architectures.
Several such scenarios are discussed in Sec.~\ref{sec:local_remote}. 
The protocol can also be extended as a software security feature to mitigate the spread of malware even if it has compromised a device's kernel. 
An antimalware application is proposed in Sec.~\ref{sec:aslp}.

\subsection{Cross-Device Authentication}
\label{sec:local_remote}

Since the CAVE possesses no information about the underlying biometrics or CAS implementation, it does not require specialized biometric sensors/behavioral modules. CAVE implementations may therefore exist on a wide array of different device types. Moreover, adding new modalities, e.g., integrating support for wearable technology sensors, does not require any change to the protocol or the CAVE architecture; only CAS client changes are required. Consequently, there is no fixed requirement on the types of CAS device hardware that may be supported, although maintaining minimal subsets of modalities might be desirable from a security perspective. 

The literature on previous variations of Vaulted Verification summarized in Sec.~\ref{sec:background} constrained discussions of authentication to a context where the transfer of sensitive information occurs between a client device and a server. Especially for mobile devices, this type of a \textit{device-server} model provides minimal security guarantees because it does not protect data cached on the device itself. 
Mobile email clients, for example, only authenticate with a server in order to download new emails. Once the emails have been downloaded, no security layer between the device and the server can protect the contents of these emails -- encrypting cached data and using the server as a trusted central authentication authority is not feasible from a usability perspective due to the loss of access when offline. Thus, in addition to device-server verification, a solution in which the user authenticates to the device itself is needed. However, if the device has been compromised by spoofing, physical theft and subsequent rooting, or by malware, then any continuous authentication layer sharing the same memory space runs the risk of compromise. Instead, a \textit{device-TCM} scheme is needed, in which the CAS and the CAVE run in two disjoint address spaces -- the CAS runs in the standard device address space and the CAVE  runs in the address space of a \textit{trusted computing module} (TCM), which by design is very difficult to compromise.

One way to accomplish address space separation is to move the CAVE into the OS kernel itself. Although this increases the \textit{cost} of the attack, this does not remove the vulnerability to malwares which compromise the kernel, e.g., DKOM rootkits~\cite{rudd2016survey}
. A better approach is to use a \textit{physically different} memory space as the TCM than that of the CPU, with a much different processing architecture, so that the CAVE resides on its own separate hardware. 
For workstations and laptops, such a memory separation could be accomplished by moving the CAVE onto a Trusted Platform Module (TPM). For smart phones, subscriber identity modules (SIM cards) could be used instead. Both TPMs and SIM cards are equipped with cryptographic modules and support permutation, indexing, and comparison operations necessary to perform CAVE functionalities. Graphical Processing Units (GPUs) also have their own memory spaces as well as programming models which are sufficiently general to support CAVE functionality, while lacking sufficient generalization to easily support most malwares. 

\begin{figure}[t]\begin{center} \includegraphics[width=\linewidth]{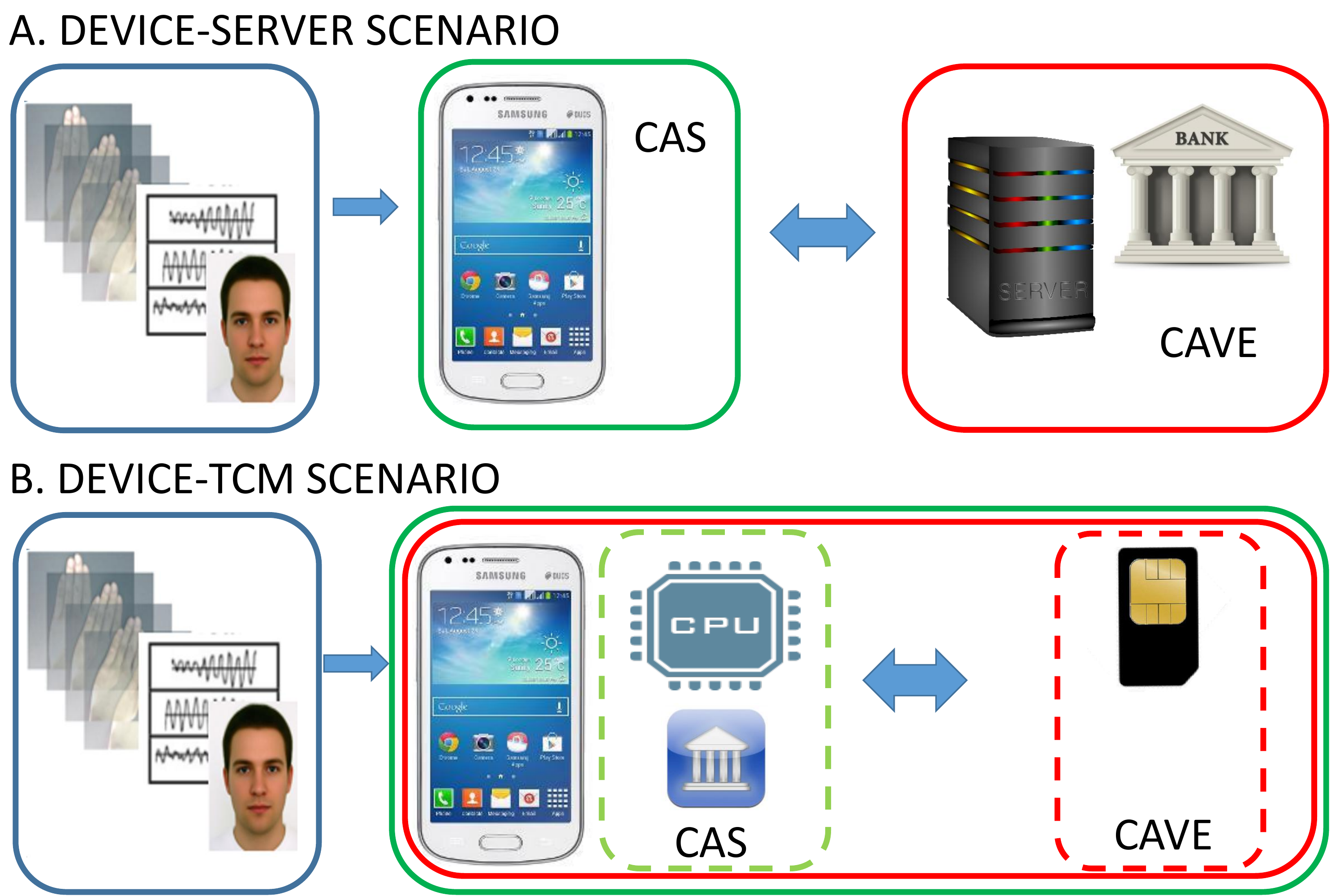}\hfil \end{center}
\caption{\small
Examples of and device-server and device-TCM instances of the CALIPER protocol. 
Solid green lines indicate CAS devices, while solid red lines indicate CAVE devices.
Dashed green and red lines indicate CAS- and CAVE-specific hardware within a device.
In the device-server scenario, the CAS layer, running on a mobile phone in this example, authenticates the user to a remote banking server's CAVE in order to process a transaction.
The CAVE runs on a remote server and challenges the CAS residing on the user's phone about the user's identity. 
In the device-TCM scenario, both the CAS and the CAVE run on the phone. The CAS runs on the same CPU as the banking application, and authenticates the user to the CAVE, which runs on a SIM card. In this scenario, the SIM card manages access to local application data by the CPU.
}
\label{fig:local_remote}
\end{figure}

Schematics of device-server and device-TCM variations of CALIPER are shown in Fig.~\ref{fig:local_remote} in the context of a mobile online banking application. Note that device-TCM and device-server applications should not be thought of as disjoint contexts since they often occur simultaneously. For example, the banking application running on a smart phone in Fig.~\ref{fig:local_remote} might leverage a CAVE running on a SIM card to verify the user, while making network connections to a bank which uses its server resident CAVE to verify the user against the same CAS. \comment{Thus, many applications of CALIPER in the mobile domain required a device-TCM implementation nested within a device-server implementation.}

\subsection{Remote Entrusting}
\label{sec:aslp}

In Sec.~\ref{sec:local_remote}, we discussed how CALIPER addresses local and remote authentication problems. 
However, CALIPER can also be applied to address more general remote entrusting issues. 
In this section, we present a novel application in which CALIPER can be applied to provide enhanced protection against malware.
\comment{, but also to serve as a digital rights management (DRM) solution.}

One method most modern operating systems use to attain robustness to attacks is to randomize the address space layout at load time. 
This technique is commonly referred to as \textit{address space layout randomization} or ASLR. 
The objective of ASLR is to make attacks more difficult by removing the attacker's ability to know the memory address space layout apriori: It is difficult to write code to hijack control flow if the address of the hook point is unknown, especially in a large 64-bit address space.
Unfortunately, the utility of ASLR assumes a trusted device kernel. 
If the device's kernel is compromised, then ASLR can be disabled or modified to yield a deterministic address space layout, e.g., by hooking the random number generator.

To this end, we introduce \textit{address space layout personalization} (ASLP), a concept which is in some ways similar to ASLR, but is far less vulnerable to kernel compromise.
\comment{
Under ASLP, an application is compiled and linked on a vendor/content distributor's server, on which the cave resides.
Randomization of code segments is keyed by vendor challenge. 
Thus, when the client attempts to load the application, the only way to correctly address segments is to “answer” the CAVE’s challenges via live biometrics or device usage samples. 
This significantly raises the bar for the attacker, since segment ordering is now user-specific and the attacker cannot deploy generic malware to hook the application by simply disabling ASLR.
}
\comment{
When distributing an application, vendors fear the types of application hooking that ASLR aims to prevent. 
Unfortunately, the security of ASLR is predicated on the assumption that the device has not been rooted, since the adversary with root access can simply disable ASLR.
}
In ASLP, compilation and linking of executables is done on the vendor's server on which the CAVE resides. 
Randomization of code segments is keyed by the CAVE's challenge. 
When the CAS client attempts to load the application, the only way to \textit{correctly address segments} is to “answer” the CAVE's challenges via live biometric/usage samples and recover the key. 
This significantly raises the bar for the attacker, since segment ordering is now user-specific and generic malware cannot hook the application by simply disabling ASLR.
If the randomization of ASLP is disabled, then the application will simply fail to load.
Of course, there is a chance that a rootkit can glean information about the load order by snooping on the address space at load time, but the address space layout for one or more particular instances is not particularly useful for a malicious code author: each user of each application has his/her own unique address space layout. Also, in the event of a security breach, the server can re-deploy code to the client, this time randomized with a different CALIPER key. 
The knowledge of a given address space layout is only applicable to a single CALIPER key for a single user.

\section{Discussion}

While it might seem grandiose to claim that the CALIPER protocol can operate using the constrained computational resources of a SIM card as a CAVE, it is important to realize that most of CALIPER's processing and storage is performed on the CAS, which we assume has at least the computational and storage capacities of modern smartphones. 
The CAVE must support primitive arithmetic and cryptographic operations as most modern SIM cards do. 
It must also be able to store the server table. 
The exact server table size is implementation dependent, but for a feasibility assessment, let us assume that we have a 128 KB SIM card for our CAVE. 
Let us further assume that we are using SHA-256 digests (32 bytes each) as our hashes, that each row in our server table has four choices, only one of which is correct, and that $K_{pr}$ and $K_{pu}$ are 2048 bit RSA keys.
Then $C(K_{pr})$, and hence $R$ take 2048 bits if we move the error correcting symbols to the CAS for compactness, which can easily be done without compromising security. 
Finally, let us assume that the server table has 4 choices per row in each of 128 rows per key.
The server then requires 256 bytes for $K_{pu}$, 32 bytes for the \uid hash, and 140 bytes ($4 \times 2$ for $R$, $4 \times 1$ for $i$, and $4 \times 32$ for $h2$) for each row in the server table.
Under the 128 KB SIM card assumption, the CAVE has enough storage capacity for the CAS to reconstruct seven 2048 bit public keys. 

Continuous authentication is in its early stages, so we acknowledge that this feasibility analysis, although it might seem reasonable at face value, may or may not be appropriate, depending on the application in question. According to the security analysis in ~\cite{johnson2014privacy} our example is \textit{secure} in comparison to conventional password authentication standards (in terms of \textit{bits of security}) for a single modality. 
Research suggests that multiple modalities serve to enhance security~\cite{sim2007continuous}, although how well they do so is still an open research question. 
As continuous authentication technologies, e.g.,~\cite{project_abacus}, are fielded, it will be interesting to see what constitutes a good balance of modalities. 
Some modalities have been documented to yield higher authentication accuracies than others, but the amount of \textit{independent information} that each modality adds, conditioned on the presence of others for particular authentication platforms has yet to be researched, as does the impact of \textit{missed detections} and missing modalities on the tradeoff between authentication performance and usability. 
The biometrics and behavioral data used in the CALIPER protocol need not be entirely available on-demand, although the extent of temporal evidence accumulation that can practically be accommodated depends on the tradeoff between security and usability. 
If the CAS consistently makes substantial errors in its guesses, then ramping up security by reducing the lengths of time windows for sensor data acquisition, and perhaps even requiring \textit{active authentication} -- i.e., explicit prompts for user input -- may be a reasonable security failover.
We leave these topics for future research.

While CALIPER adds a significant improvement to the security of all proposed and fielded continuous authentication technologies that we are aware of, it does not eliminate all security vulnerabilities. 
As with any biometric/biocryptographic protocol, the security of the protocol is compromised if the biometrics are compromised at the time of enrollment.
If the device is root-compromised after enrollment, it is still extremely difficult for an attacker to access local or remote data while the device is not in use.
When the device is used while in a root-compromised state, however, no biometric-based system can protect users in the long run without a re-keyed protocol because an attacker can eventually gather the data to construct a complete biometric profile of the user. 
While this is not required to recover $K_{pr}$ alone, the vulnerability of $K_{pr}$ is partially ameliorated by challenge-response: For multiple attempts, the key alone is insufficient and the CAVE would still be able to reject the attacker.
Another option would be to move generation and retrieval of $K_{pr}$ to the CAVE. 
This protects $K_{pr}$ from immediate compromise and potential key-inversion attacks by malware on the CAS, but eventually, after multiple legitimate authentications on behalf of the user, a malware could intercept raw sensor data to develop a more complete biometric profile and thus answer the challenges.
Thus, as a measure of intrusion detection, it is critical that a CAVE occasionally send a challenge that is expected to fail and check that no valid key is returned before the attacker has a chance to compromise the biometric/behavioral data.
A compromised kernel is generally non-trivial for any authentication system. This is one of the reasons why we extended the CALIPER protocol as a remote entrusting mechanism to help protect the CAS's kernel (cf. Sec.~\ref{sec:aslp}).

\comment{
Finally, we mention that assuming generation of $K_{pr}$, $K_{pu}$, $R$, and the CAVE's permutation key are indistinguishable under a chosen ciphertext attack (IND-CCA2), then CALIPER is IND-CCA2 secure, which also implies non-malleability, indistinguishability under a chosen ciphertext attack (IND-CCA), and indistinguishability under a chosen plaintext attack (IND-CPA).
This only says that under chosen plaintext and ciphertext pairs fed to encryption/decryption oracles, we have no way to correlate ciphertexts to different plaintexts or different ciphertexts to different ciphertexts -- it says nothing about the number of ``bits of security'' which is dependent on implementation, modality, and number of error correcting symbols. Also, what constitutes a particular ``plaintext'' is the domain of the hypothesis space across all classifiers that maps to a particular range, so reasonable false accept rate (FAR) assumptions must be met in order to provide a sufficiently large keyspace.
}

{\small
\bibliographystyle{ieee}
\bibliography{caliper}

\begin{thebibliography}{10}\itemsep=-1pt

\bibitem{project_abacus}
Project abacus is an atap project aimed at killing the password, 2015.

\bibitem{alzahrani2014remote}
H.~Alzahrani and T.~E. Boult.
\newblock Remote authentication using vaulted fingerprint verification.
\newblock In {\em SPIE Defense+ Security}, pages 90750K--90750K. International
  Society for Optics and Photonics, 2014.

\bibitem{aussel2009smart}
J.-D. Aussel, J.~d’Annoville, L.~Castillo, S.~Durand, T.~Fabre, K.~Lu, and
  A.~Ali.
\newblock Smart cards and remote entrusting.
\newblock In {\em Future of Trust in Computing}, pages 38--45. Springer, 2009.

\bibitem{azzini2008fuzzy}
A.~Azzini, S.~Marrara, R.~Sassi, and F.~Scotti.
\newblock A fuzzy approach to multimodal biometric continuous authentication.
\newblock {\em Fuzzy Optimization and Decision Making}, 7(3):243--256, 2008.

\bibitem{ceccato2008remote}
M.~Ceccato, Y.~Ofek, and P.~Tonella.
\newblock Remote entrusting by run-time software authentication.
\newblock In {\em SOFSEM 2008: Theory and Practice of Computer Science}, pages
  83--97. Springer, 2008.

\bibitem{dodis2004fuzzy}
Y.~Dodis, L.~Reyzin, and A.~Smith.
\newblock Fuzzy extractors: How to generate strong keys from biometrics and
  other noisy data.
\newblock In {\em Advances in cryptology-Eurocrypt 2004}, pages 523--540.
  Springer, 2004.

\bibitem{feng2012continuous}
T.~Feng, Z.~Liu, K.-A. Kwon, W.~Shi, B.~Carbunar, Y.~Jiang, and N.~K. Nguyen.
\newblock Continuous mobile authentication using touchscreen gestures.
\newblock In {\em Homeland Security (HST), 2012 IEEE Conference on Technologies
  for}, pages 451--456. IEEE, 2012.

\bibitem{frank2013touchalytics}
M.~Frank, R.~Biedert, E.-D. Ma, I.~Martinovic, and D.~Song.
\newblock Touchalytics: On the applicability of touchscreen input as a
  behavioral biometric for continuous authentication.
\newblock {\em Information Forensics and Security, IEEE Transactions on},
  8(1):136--148, 2013.

\bibitem{guennoun2009continuous}
M.~Guennoun, N.~Abbad, J.~Talom, S.~M.~M. Rahman, and K.~El-Khatib.
\newblock Continuous authentication by electrocardiogram data.
\newblock In {\em Science and Technology for Humanity (TIC-STH), 2009 IEEE
  Toronto international conference}, pages 40--42. IEEE, 2009.

\bibitem{johnson2013vaulted}
R.~Johnson and T.~E. Boult.
\newblock With vaulted voice verification my voice is my key.
\newblock In {\em Technologies for Homeland Security (HST), 2013 IEEE
  International Conference on}, pages 453--459. IEEE, 2013.

\bibitem{johnson2013voice}
R.~Johnson, T.~E. Boult, and W.~J. Scheirer.
\newblock Voice authentication using short phrases: Examining accuracy,
  security and privacy issues.
\newblock In {\em Biometrics: Theory, Applications and Systems (BTAS), 2013
  IEEE Sixth International Conference on}, pages 1--8. IEEE, 2013.

\bibitem{johnson2013secure}
R.~Johnson, W.~J. Scheirer, and T.~E. Boult.
\newblock Secure voice-based authentication for mobile devices: vaulted voice
  verification.
\newblock In {\em SPIE Defense, Security, and Sensing}, pages 87120P--87120P.
  International Society for Optics and Photonics, 2013.

\bibitem{johnson2014privacy}
R.~C. Johnson.
\newblock {\em Privacy enhanced remote voice verification}.
\newblock University of Colorado at Colorado Springs, 2014.

\bibitem{jorgensen2011mouse}
Z.~Jorgensen and T.~Yu.
\newblock On mouse dynamics as a behavioral biometric for authentication.
\newblock In {\em Proceedings of the 6th ACM Symposium on Information, Computer
  and Communications Security}, pages 476--482. ACM, 2011.

\bibitem{juels2006fuzzy}
A.~Juels and M.~Sudan.
\newblock A fuzzy vault scheme.
\newblock {\em Designs, Codes and Cryptography}, 38(2):237--257, 2006.

\bibitem{klosterman2000secure}
A.~J. Klosterman and G.~R. Ganger.
\newblock Secure continuous biometric-enhanced authentication.
\newblock 2000.

\bibitem{leggett1991dynamic}
J.~Leggett, G.~Williams, M.~Usnick, and M.~Longnecker.
\newblock Dynamic identity verification via keystroke characteristics.
\newblock {\em International Journal of Man-Machine Studies}, 35(6):859--870,
  1991.

\bibitem{liu2009optimal}
J.~Liu, F.~R. Yu, C.-H. Lung, and H.~Tang.
\newblock Optimal combined intrusion detection and biometric-based continuous
  authentication in high security mobile ad hoc networks.
\newblock {\em Wireless Communications, IEEE Transactions on}, 8(2):806--815,
  2009.

\bibitem{rudd2016survey}
E.~Rudd, A.~Rozsa, M.~Gunther, and T.~Boult.
\newblock A survey of stealth malware: Attacks, mitigation measures, and steps
  toward autonomous open world solutions.
\newblock {\em arXiv preprint arXiv:1603.06028}, 2016.

\bibitem{scandariato2008application}
R.~Scandariato, Y.~Ofek, P.~Falcarin, and M.~Baldi.
\newblock Application-oriented trust in distributed computing.
\newblock In {\em Availability, Reliability and Security, 2008. ARES 08. Third
  International Conference on}, pages 434--439. IEEE, 2008.

\bibitem{scheirer2013beyond}
W.~J. Scheirer, W.~Bishop, and T.~E. Boult.
\newblock Beyond pki: The biocryptographic key infrastructure.
\newblock In {\em Security and Privacy in Biometrics}, pages 45--68. Springer,
  2013.

\bibitem{scheirer2007cracking}
W.~J. Scheirer and T.~E. Boult.
\newblock Cracking fuzzy vaults and biometric encryption.
\newblock In {\em Biometrics Symposium, 2007}, pages 1--6. IEEE, 2007.

\bibitem{shepherd1995continuous}
S.~Shepherd.
\newblock Continuous authentication by analysis of keyboard typing
  characteristics.
\newblock 1995.

\bibitem{sim2007continuous}
T.~Sim, S.~Zhang, R.~Janakiraman, and S.~Kumar.
\newblock Continuous verification using multimodal biometrics.
\newblock {\em Pattern Analysis and Machine Intelligence, IEEE Transactions
  on}, 29(4):687--700, 2007.

\bibitem{wilber2012secure}
M.~Wilber, T.~E. Boult, et~al.
\newblock Secure remote matching with privacy: Scrambled support vector vaulted
  verification (s 2 v 3).
\newblock In {\em Applications of Computer Vision (WACV), 2012 IEEE Workshop
  on}, pages 169--176. IEEE, 2012.

\bibitem{wilber2012privv}
M.~J. Wilber, W.~J. Scheirer, and T.~E. Boult.
\newblock Privv: Private remote iris-authentication with vaulted verification.
\newblock In {\em Computer Vision and Pattern Recognition Workshops (CVPRW),
  2012 IEEE Computer Society Conference on}, pages 97--104. IEEE, 2012.

\end{thebibliography}
}

\end{document}